# Superconductivity at T≈200 K in Bismuth Cuprates Synthesized Using Solar Energy

J. G. Chigvinadze (1), J. V. Acrivos (2) , S. M. Ashimov (1), D. D. Gulamova (3), G. J. Donadze (1)

((1) I.Javakhishvili Tbilisi State University, E.Andronikashvili Institute of Physics, Tbilisi, Georgia, (2) San Jose State University, San Jose, CA, USA, (3) The Institute of Materials Science SPA "Physics-San" of Academy of Science, Tashkent, Uzbekistan)

## Abstract

When investigating low-frequency (0.1 Hz) oscillations of multiphase high-temperature cuprate superconductors (HTCS) $Bi_{1,7}Pb_{0,3}Sr_2Ca_{(n-1)}Cu_nO_y$ (n=2-30), a wide attenuation peak ($\Delta T \sim 100$ K) with a maximum at T≈200 K was detected. This peak was particularly pronounced in field cooling (FC) experiments, i.e. after abrupt cooling of the sample in the external magnetic field at the temperature $T<T_c$ with subsequent slow warming up to room temperature with invariance of the applied field. The attenuation peak height depended on the preliminary orientation (before cooling) of the samples θ in the measured permanent magnetic field H. On the one hand, it is well known that, after the FC procedure and subsequent slow warming up, at the temperatures close to the critical temperature $T_c$, the attenuation peak associated with "melting" of the Abrikosov frozen vortex structure and its disappearance at $T > T_c$ is detected in monophase samples. At the same time, in most multiphase bismuth HTCS samples, synthesized using solar energy and superfast quenching of the melt, the attenuation peak with the maximum at T≈200 K was observed.

Depending on the conditions of synthesis, the attenuation peak could be two-humped and could be located in the temperature range much wider than $T_c$ of the major superconducting phase. We assume that this is due to the existence of frozen magnetic fluxes (after FC) in superconducting "dropping" regions, which gradually (with increasing temperature) transfer into the normal state and release pinned vortex threads. This fact could be a cause of observed dissipative processes, so as also the evidence of the existence of superconductivity at T ≥240 K.

## 1. Introduction

Despite a lot of data have been accumulated since discovery of cuprate high-temperature superconductors in 1986, there have been few attempts to perform microscopic studies of how a superconducting state emerges in such materials. The question of why some substances become superconducting at relatively high temperature led to the investigation of various properties and characteristics of high-temperature superconductors (HTS) both in superconducting and normal states. Based on firmly established experimental facts, Nobel Prize Laureate A.J. Leggett [1] formulated the following requirement (along with other ones) to the modern theory of superconductivity: to explain why just in cuprates ($CuO_2$) high-temperature superconductivity is observed. Moreover, the theory must define what the room-temperature superconductor (RTS) is. For today, a multielectron theory of superconductivity satisfying these requirements has been elaborated using an artificial intellect [2].

About 50 years ago, N. Mott introduced the term "pseudogap" [3]. He used it to indicate a minimum in the density of electron states at the Fermi level resulting from Coulomb repulsion between electrons at lattice nodes or being a precursor of formation of a forbidden gap in disordered systems, or either being a consequence of a combination of these two factors. Later it was noticed that the pseudogap could be a sign of the existence of noncoherent-in-phase Cooper pairs in the superconductor at $T>T_c$. After the discovery of the pseudogap in cuprate HTS, many considered exactly this mechanism of its formation to be the most probable, which was indirectly confirmed by fluctuation superconductivity observed in HTS at temperatures much higher than $T_c$. The region of superconducting fluctuations is quite large and experimentally can reach several tens degrees [4]. At the same time, for antiferromagnetic (AFM) fluctuations characterized by an order of greater energy, the existence of a critical area hundreds degrees wide does not seem so incredible. There was a point of view that the pseudogap phase did not precede the superconducting phase, but competed with it. Though, in essence the question remained open.

In work [5], discussing the causes of emergence of the pseudogap at relatively high temperature of about 100-150 K, a big group of scientists arrived to a conclusion that it was a completely special state of substance having no relation to superconductivity. While not giving the answer to the question on the pseudogap origin, the authors nevertheless used three methods to investigate HTS Bi/Pb (2201). The first one was the most widely used method of detection of the pseudogap – angle-resolved photoemission spectroscopy (ARPES), which undoubtedly is an effective technique for a certain range of problems, however it gives only a superficial (in the literal sense) idea of a substance. The second method was time-resolved spectroscopy. The third one involved the measurement of the magneto-optic Kerr effect, which allowed detecting a change in the magnetic order inside the crystal at T*=132 K and identifying it as a phase transition. Hence, according to the authors, for the HTS crystal to begin to superconduct, it was necessary that it underwent two phase transitions as the temperature decreased, first the emergence of the pseudogap and then of superconductivity. On the other hand, it was assumed that the origin of the pseudogap and superconductivity could have common features. This fact explains a great number of reviews, e.g. [6], and articles devoted to this issue [7-9].

At the same time, in [10] a survey of theoretical and experimental data pointing out the possibility of spontaneous phase separation in a number of HTS systems (especially in the area of underdoped compositions) is given. This separation occurs on a microscopic scale so that the system divides into metallic (superconducting) and dielectric (antiferromagnetic) domains. In [11] it was shown that the existence of superconducting "drops" in many cases explains anomalous diamagnetism over the temperature range higher than $T_c$. On the other hand, depending on external effects, phase transitions and variations in phase stability can be observed. For instance, a phase stable under certain temperature conditions becomes instable when the temperature increases or decreases.

The results obtained in [12] indicate that maybe both are right and that two different states coexist in the pseudogap phase: in one of them (at $T_c<T<T_{pair}$) there are noncoherent Cooper pairs, while the other is realized over a wider temperature range (at $T_c<T<T^*$, where $T^*>T_{pair}$ – the temperature of emergence of the pseudogap) and is of a nonsuperconducting nature. The authors [12] arrived to this conclusion when studying the ARPES-spectra of single crystals Bi-2212 and Bi-2201 with different levels of doping and analyzing temperature dependences of the shapes of spectral lines measured in the anti-node direction of the Brillouin zone. This served as a basis for the authors to make a hypothesis about the formation of another pseudogap state at $T=T_{pair}$, which at $T<T_{pair}$ coexisted with the one formed at $T_{pair}<T<T^*$ and which was supposedly caused by formation of noncoherent Cooper pairs.

For today it is considered established [6,7,9,13] that the electron-phonon coupling in cuprate superconductors is quite strong. Under such conditions, any change in the electron structure must inevitably lead to the changes in phonon characteristics and other physical properties. Thus, for instance, when studying the ultrasonic velocity [14] in superconductor $GdBaSrCu_3O_{7-x}$ ($T_c$ = 82 K), there was observed a step-like anomaly at $T_g$=245 K ($T_g\approx 3T_c$) clearly indicating a change in lattice rigidity. It is well known that HTS manifest anomalous behavior over the temperature range 150-250 K. Thus, in works [15,16] it was shown that, in systems Bi-2201 and Bi-2212 at $T\approx 240$ K, the coefficient of thermal expansion (CTE) changed the sign with cooling and became negative. The values of lattice parameters increase, i.e. the effect is volumetric.

The authors of works [17,18] believe that the emergence of superconductivity in cuprate superconductors $(R)Ba_2Cu_3O_{6+x}$ (R =Y, Gd, Tm, Ho) at relatively high temperature is expressed in the form of a temperature "echo". Such an echo is characteristic of superconductors with strong coupling, and determined by anomalies of the heat capacity [18]. The mechanism of such nucleation is assumed to be related to the occurrence of the order parameter amplitude with the absence of phase coherence of paired current carriers [19]. In works [20,21] they introduced the notion of emergence of two pseudogaps, the first of which was associated with the crossover phenomenon unrelated to superconductivity and another – with the phenomenon of phase transition caused by superconductivity. Due to currently unidentified causes, the density of quasiparticle states at the Fermi level begins to decrease [22-24].

We are most impressed by the point of view and the experimental data of the authors of works [17,18] who presented the results of the analysis of the experimental heat capacity over a wide range of normal and superconducting states. An anomaly $T_h$ steadily manifested in the interval 250–290 K was observed for all samples. The anomaly $T_h$ in appearance resembles the phenomenon related to the phase transition. The interconnection between the $T_h$ and $T_c$ processes prompts that the phenomena at $T_c$ and $T_h$ are affiliated and, according to the authors, they both are related to the superconductivity.

Another example are the results of measurements of the coefficient of heat expansion $\alpha(T)$ of single-crystals $YBa_2Cu_3O_{6.95}$ and $YBa_2Cu_3O_7$ carried out over the temperature range 5-500K for three orthorhombic axes [25]. Depending on $\alpha(T)$, for the sample $YBa_2Cu_3O_{6.95}$, the anomalies are observed as at the temperature of the superconducting transition $T_c$=93 K, so also at $T_g$=280 K ($T_g \approx 3T_c$).

Practically in all abovementioned works, HTS with a single major superconducting phase were studied. In this work multiphase cuprate semiconductors synthesized using the solar energy are considered, and the results of the investigation of superconducting phases of the samples Bi/Pb (2:2:2:3), (2:2:4:5), (2:2:19:20) and (2:2:29:30), which, along with low-temperature phases with $T_c \approx 107$ K, contain other higher-temperature superconducting phases, are presented. In our investigation, percentage of these phases varied from 1% to 98% depending on the task set when searching for optimal thermodynamic conditions of synthesis of HTS at specified temperature and time with the aim of obtaining the superconducting phases with maximum $T_c$.

## 2. Multiphase samples under study

Using the recently developed technology Solar Fast Alloys Quenching-T (SFAQ-T) [26,27], based on glass-crystal and X-ray amorphous precursors, we obtained decomposition-resistant textured superconducting systems $Bi_{1.7}Pb_{0.3}Sr_2Ca_{(n-1)}Cu_nO_{10-y}$ (n=2-30) with critical temperature of the superconducting transitions up to $T_c \geq 181$ K [28]. We synthesized the precursors by quenching of the melt which was produced by heating with solar radiation. The samples under study were axially symmetric pressed tablets. The appearance of precursors is shown in Fig. 1a,b,c and nanosize nuclei on the precursor-plate surface are presented in Fig. 1d.

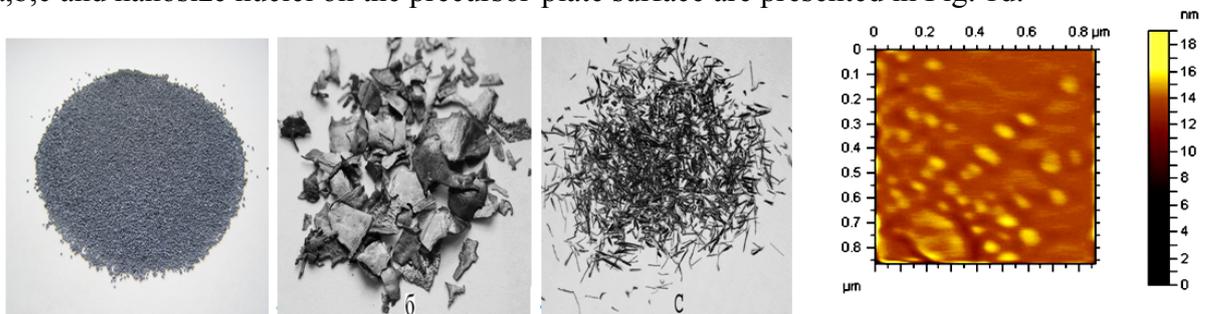

**Fig. 1. Precursors of nominal compositions $Bi_{1.7}Pb_{0.3}Sr_2Ca_{(n-1)}Cu_nO_y$ (n=2-30), obtained by quenching of the melt under concentrated solar radiation: a – spherulites; b – plates and pieces; c – needles; d – nanosize nuclei in glass-crystal precursors.**

The investigation of the phase composition of precursors showed the following. On the diffractograms of precursors with n=3-12 the halo and separate reflexes on its background indicated their glass-crystal state. An increase in n≥12 and the corresponding increase in the concentration of CaO and CuO, led to destabilization of the glass phase. Initiation of precursor reflexes of nominal n=3-12 showed that, with quenching of the melt, the nuclei with crystallographic planes were frozen for the superconducting phase (2201) - [113, 115, 019, 024, 028, 0113], phase (2212) – [017, 0111, 0210, 0010], and phase (2223) - [115, 119, 0015], respectively. These planes could be the nuclei on which it is possible at thermal treatment the completion of crystal lattice of that superconducting phase for which the thermodynamic conditions of formation were optimal at specified temperature and time. In compositions with n=5-20, 25, 30, at 100-700°C, new phase X1 with the structure differing from superconducting phases (2201), (2212), (2223), (2234) was formed. Another new phase X2 was formed when the temperature increased up to 846°C. The time of heat treatment (3, 24 and 48 hours) did not affect the stability and changes of the crystal structure of X1 and X2 phases. According to diffractograms, the intensity of the reflexes of superconducting phase X2 increased in a series with increasing n. The parameters of a unit cell of phase X2 were a=3.9015Å; b=3.8185Å; c=135.5947Å.

## 3. Methods of investigation

In this work, to determine the critical temperature of the superconducting transition $T_c$, we used a magnetomechanical method of torsion oscillations realized using an automated multipurpose device [29], having the sensitivity comparable with that of a SQUID magnetometer [30], designed at E.Andronikashvili Institute of Physics (Tbilisi, Georgia). The investigation was carried out at low-frequency axially-torsion oscillations (0.1÷1 Hz) in a permanent magnetic field with strength H and showed a significant effect of the background of the experiment, the value of H, initial orientation of the sample and the direction of variation in the temperature of the sample (cooling or warming) on the obtained results.

The torsion instrumentation used is especially sensitive to reorientation of magnetic moments of the materials under study in external magnetic fields. As all studied HTS possess own magnetic moments, the experiments of this kind are quite informative when studying structural transitions, especially when such transitions are accompanied by reorientation of magnetic moments including the normal state at $T>T_c$.

The method of torsion oscillations was first used for investigation of energy loss (dissipation) in the mixed state of hard superconductors in works [31,32], where quite high sensitivity of the torsion system ($10^{-17}$W) was shown. The use of these possibilities allowed determining the critical parameters such as $T_c$ or the first critical field $H_{c1}$ [33-36], studying the anisotropy of pinning force $F_p$ in high-temperature oxide superconductors [37,38], and also the intrinsic magnetic characteristics of HTS samples [29,39-42]. The studies of this kind allow investigating the issue of order parameter symmetry, judging about the mechanism of pairing and hence about the mechanism of high-temperature superconductivity. Besides, studying the dissipative processes in the vicinity of $T_c$, we can observe and investigate the effects of melting of the Abrikosov magnetic vortex lattice in HTS [43-45].

The phase transition temperature $T_c$ was determined not only by the frequency $\omega=2\pi/t$ of the superconductor oscillating in a permanent magnetic field H, but also by the character of the dependence of the dissipative process δ(T), where δ is the logarithmic decrement of attenuation of oscillations. These two characteristics t(T) and δ(T), being measured in parallel and complementing each other, provide information on the presence or absence of magnetic vortex threads in the sample under study, which allows judging the state of the sample.

In case there are no magnetic moments in the sample, dissipation and frequency of oscillations do not depend on the external magnetic field. For instance, the superconducting sample oscillates in magnetic fields $H<H_c$, or the internal moments are zero or either are disoriented or unfixed. The presence of pinned magnetic dipoles generates a nonzero magnetic moment M in the sample even at room temperature. The interaction between M and H generates the moment $\tau = M \times H \sin\alpha$, where $\alpha$ is the angle between M and H. This additional moment $\tau$ affects both the immobile and oscillating systems, making the dissipation and frequency of oscillations dependent on the external magnetic field, especially with the presence of vortex threads in the mixed ($H > H_{c1}$) state of separate areas of HTS under study.

As was shown in work [46], for the superconductors in the mixed state, the interaction between pinned and unpinned (free) Abrikosov vortices plays an important role in the dynamic oscillating processes affecting both the frequency $\omega$ and the dissipation $\delta$ of the energy of oscillations. It is well known that the pinning force also depends much on the temperature, e.g. it tends to zero approaching $T_c$. At the same time, the concentration of free vortices increases, while the value of $\omega$ sharply decreases. The results of corresponding investigation of the monophase sample Bi/Pb (2-2-2-3), synthesized by a standard solid-state reaction, shown in Fig, 2, can be compared with the results of investigation of multiphase samples. It is evident that, as the temperature increases approaching $T_c$ a sharp attenuation maximum appears in the $\delta(T)$ dependence. This maximum is related to the decrease in the pinning force and gradual liberation of Abrikosov vortices from pinning centers and their viscous motion in the sample matrix with oscillations. Approaching $T_c$, this process is replaced with the process of melting of the Abrikosov vortex lattice with its gradual disappearance at $T=T_c$, which leads to an abrupt increase in the value of t and an similar abrupt reduction of attenuation of $\delta(T)$ in the immediate vicinity of $T \sim T_c$.

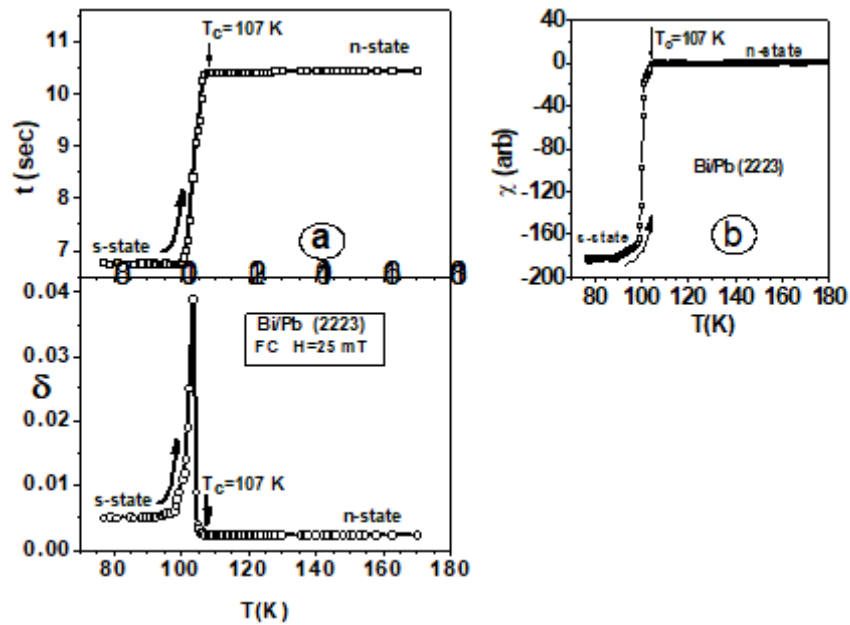

**Fig. 2. a) – Temperature dependence of the logarithmic decrement of attenuation $\delta$ and the oscillation period t for the monophase sample Bi/Pb (2-2-2-3), synthesized by a standard solid-state reaction and studied after abrupt cooling to T=77K in the magnetic field H=25mT; b) – magnetic susceptibility.**

In most cases, the investigations were accompanied by testing of the samples by measuring the electric resistance (4-contact method) and magnetic susceptibility. The characters of dependences $\chi(T)$ and $t(T)$ coincide, which is evident from Fig. 2a,b ($T_c$=107 K). Measurements were carried out with increasing temperature from T=77 K.

It should be noted that, in all monophase Bi/Pb samples with temperature above Tc, there were not observed other superconducting phases with higher temperature. At the same time, in the majority of samples of multiphase bismuth HTS, synthesized using solar energy and superfast quenching of the melt, we detected an attenuation peak or peaks with a maximum at T $\geq$ 200 K, which is in the significantly larger region than $T_c$ of the major superconducting phase.

Figure 3 shows the results of investigation of one of the samples Bi/Pb (2245) fabricated under the conditions of the technological process close to nominal. There are shown the temperature dependences of period t and the decrement of attenuation of oscillations $\delta$ under field cooling (FC) conditions in magnetic field H=150mT. In the figure, the critical temperatures for phases (2212) with $T_c \approx 80K$ and (2223) with $T_c \approx 100K$ are taken in a circle. Especially notable is the transition from $T_1$=150 K to $T_2$=236 K. This transition is accompanied by a wide attenuation peak of $\delta$ at T≈200 K.

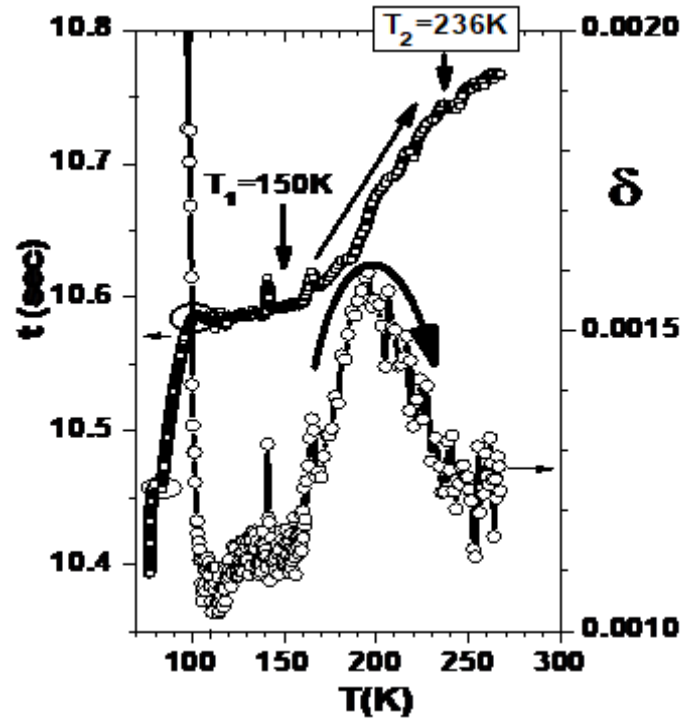

**Fig. 3. Temperature dependence of the period t and the decrement of attenuation of oscillations $\delta$ of the sample Bi/Pb (2:2:4:5) after abrupt FC in the magnetic field H=150mT to T=77 К.**

As in the case of a monophase superconductor shown in Fig. 2, here the attenuation peak indicates that it accompanies a wide superconducting transition up to T ≈ 240 K. It is obvious that this peak is related to the Abrikosov vortices frozen after the FC procedure and pinned throughout the volume of the multiphase superconductor. These vortices liberate with increasing temperature and decreasing pinning force, respectively, and hence, with their viscous motion in a sample gradually increase the value of $\delta$ up to T≈200 K. A further increase in T leads to a decrease in $\delta$ due to a gradual approach to $T_n$. As it takes place, the vortex lattice melts [43, 44] and disappears altogether when the sample transfers into a normal state. In some cases, $T_n \approx$ 240-255 K.

The character of dependences $\delta(T)$ and $t(T)$, and, respectively the values of $T_c$ of the superconducting phases of the studied multiphase sample essentially depend on the parameters of the technological process. For example, in Fig. 4 it is shown the results for two superconducting systems Bi/Pb (2:2:19:20), identical in composition but with different temperature and duration of annealing.

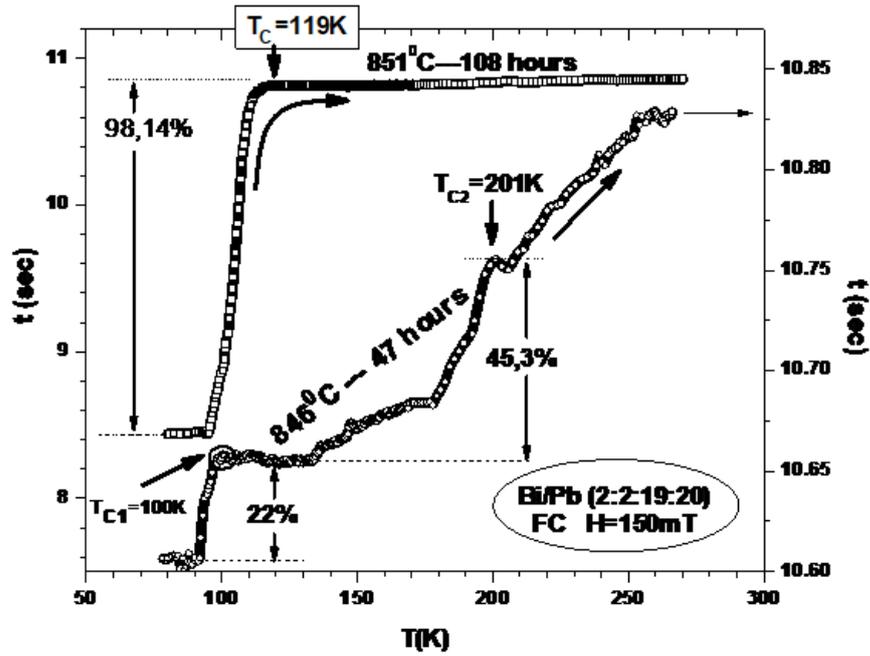

**Fig. 4.** Temperature dependences of the period of oscillations t of the samples Bi/Pb (2:2:19:20) after abrupt FC to T=77K in the magnetic field H=150mT. The temperature and time of annealing are given in the corresponding curves.

Fig. 4 shows purely subjective estimates of the contribution (in percentage) of superconducting phases to the measured dependence t(T) of both samples. As we see, the high temperature of annealing (851°C) led to the dependence with $T_C$ =119 K, which in character is practically identical to that for the monophase HTS (98.14% of the major superconducting phase). As in the case of really monophase superconductors, this transition is also accompanied by a characteristic dependence δ(T) with the attenuation peak, which is shown in curve 2 in Fig. 5 (compare with Fig. 2a).

The annealing at 846°C essentially changed both dependences t(T) and δ(T). As we see (Fig. 4), critical temperatures $T_{C1}$=100 K (22%) and $T_{C2}$ =201 K (contribution 45.3%).are particularly clearly manifested in the dependence t(T). It is noteworthy that the dependence δ(T) is two-humped (two attenuation peaks) in the temperature region T ≥ 200 K. As the attenuation peaks are associated with the Abrikosov vortices and their viscous motion along the matrix of the sample, the existence of the second peak allowed us to assume that there was a superconducting transition of another phase with temperature higher than $T_{C2}$ =201 K (curve 1 in Fig. 5). If the attenuation peak (maximum) related to $T_{C2}$ is located at T=201 K, then the second peak located at T=220 K (higher than $T_{C2}$) justifies this assumption. We managed to detect this high-temperature phase in the experiments with modified conditions of the FC procedure. Particularly, longer exposure of the HTS under study at T=77 K after its abrupt cooling in the magnetic field H allowed the critical temperature $T_{C3}$ ≈ 240 K to manifest itself. It should be noted that we observed the two-humped nature of the dependence t(T) earlier with the sample Bi/Pb (2:2:4:5) with $T_C$ =181 K, where the second attenuation peak also indicated the superconducting transition with $T_n$ ≈ 230 K [28].

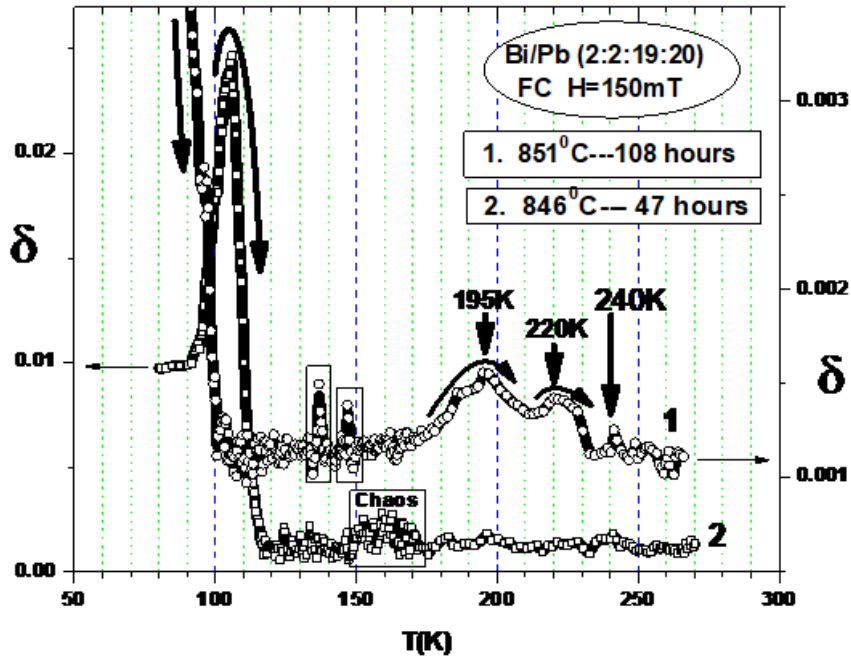

**Fig. 5.** Temperature dependences of the decrement of the attenuation of oscillations δ measured simultaneously with the dependences shown in Fig. 4 for multiphase samples Bi/Pb (2:2:19:20) after abrupt cooling to T=77 К in the magnetic field:
1) – after annealing at T= 851°C for 108 hours;
2) - after annealing at T=846°C for 47 hours.

Thus, complementing each other, the experimental results for dependences t(T) и δ(T), shown in Figs. 4 and 5, allow us to make a conclusion about the existence of the frozen Abrikosov magnetic vortex structure in multiphase HTS after the FC procedure and hence about the existence of superconductivity at relatively high temperature T ≥240 K.

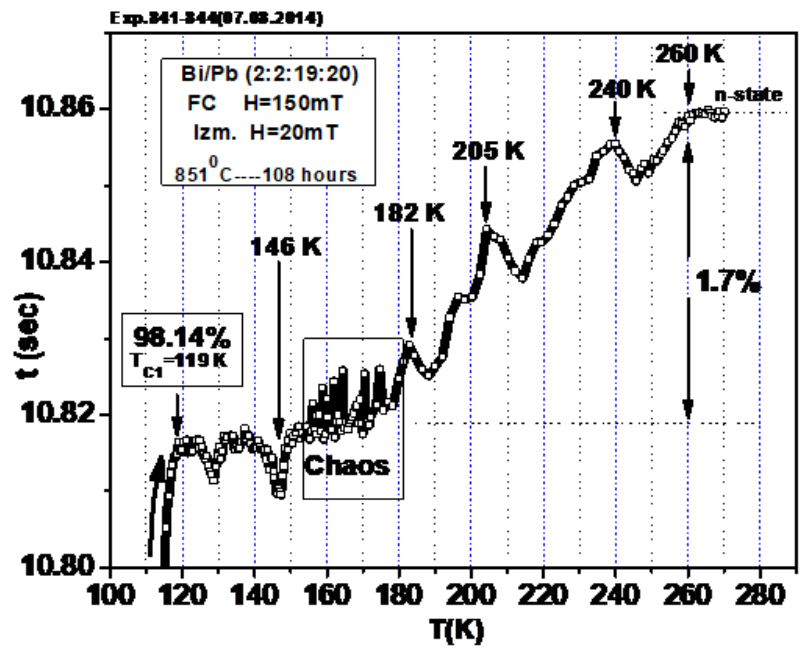

**Fig. 6.** Temperature dependences of the period of oscillations t of the sample Bi/Pb (2:2:19:20) after annealing at T= 851°C for 108 hours. Field cooling to T=77K in the magnetic field H=150 mT.

Special attention should be paid to that, for the sample with $T_c$=119 K (annealing at 851°C for 108 hours) shown in Fig. 4, the variation of the period of oscillations over the temperature range from 77K to 300K makes up a few seconds, while, for the sample with annealing at 846°C for 47 hours, this variation is much less – tenths of a second. It is noteworthy that that the variation of the period t in seconds is inherent in hard superconductors with quite strong pinning of the Abrikosov vortices. In other words, the stronger is the pinning, the harder is the oscillating system with the sample, and hence the smaller is the value of t (in seconds). As mentioned above, the methods used in this work have the sensitivity comparable with that of the SQUID magnetometer [30]. Thus, for example, the period of oscillations is determined with the accuracy to the fourth sign after the comma ($10^{-4}$ s). Thus, the dependence t(T) (Fig. 4) for the sample with $T_c$=119 K on the second scale, where the contribution of the major superconducting phase to the variation of the period from 77K to 280K makes up 98.14%, does not give visually complete information about the processes proceeding in the HTS under study at the temperature above $T_c$=119 K. In this connection, in Fig. 6 it is shown the dependence t(T) in the interval from the critical temperature $T_c$ of main phase up to T=260-270 K, at which the sample transfers to the n-state. This state is determined by the period of oscillations having reached the value that the HTS under study had in the given magnetic field before cooling.

In the dependence shown in Fig. 6, we can see the specific features at temperatures 130, 146, 182, 205 and 240 K, which is surely related to the presence of a set of superconducting phases in the given multiphase sample Bi/Pb (2:2:19:20) after annealing at T= $851^0$C for 108 часов, although their contribution to the dependence t(T) makes up only 1.7%.

We should emphasize the fact that, in this region, the value of the variation of the period of oscillations makes up hundredths of a second ($10^{-2}$ s). This fact indicates not only the weaker pinning in comparison with the sample after annealing at T=$846^0$C for 47 hours (Fig. 4), but also the high sensitivity of the torsion system even to slight variations in the frozen Abrikosov vortices associated with high-temperature superconducting phases.

Besides, the experimental result (Fig. 6) shows that homologous nuclei formed during annealing of the sample, which are the crystallization centers for HTS phases, are preserved even at high temperature of annealing ($851^0$C), although their contribution to the measurements is small.

## Conclusion

In the low-frequency dynamic experiments in the course of investigation of the magnetic properties of multiphase cuprate superconductors $Bi_{1,7}Pb_{0,3}Sr_2Ca_{n-1}Cu_nO_y$ (n=2-30), synthesized using solar energy and superfast quenching of the melt, a wide peak (ΔT~100 K) with the maximum at T≈200 K was detected. In some cases, there were detected two attenuation peaks associated with the processes in the vicinity of the superconducting phases with higher temperature. The analysis of the nature of obtained dependences and their comparison with other available results associated with the processes in the vicinity of critical temperature $T_c$, allows us to infer that, up to T≈245-250 K, there exists a magnetic vortex structure frozen after the FC cooling and hence about the existence of superconductivity in the multiphase cuprates under study.